\documentclass{iopart} 
\usepackage{graphicx} 
 
\begin{document}

\title{Towards half-metallic interfaces: the Co$_2$CrAl/InP contacts}
\author{Iosif Galanakis 
\footnote[3]{To whom correspondence should be addressed, e-mail: 
I.Galanakis@fz-juelich.de}}
 
\address{Institut f\"ur Festk\"orperforschung, 
Forschungszentrum J\"ulich, D-52425 J\"ulich, Germany}

\begin{abstract}
Although the interest on half-metallic Heusler alloys, susceptible to be used in 
spintronic applications, has considerably grown, their interfaces with semiconductors 
show very low spin-polarization. I identify  mechanisms which can keep the high
spin-polarization at the interface (more than 80\% of the 
electrons at the Fermi level are 
of majority spin)  although the half-metallicity is lost. The large enhancement of the 
Cr moment at the interface between a  CrAl terminated Co$_2$CrAl(001) spacer
and the InP(001) semiconductor weakens the effect of the  interface states resulting
in this high spin-polarization.
On the other hand the Co$_2$CrAl/InP interfaces made up by a Co layer and either an In 
or a P one  show a severe decrease of the Co spin moment but Cr in the subinterface
layer is  bulklike and the resulting spin-polarization is similar to the 
CrAl-based interfaces.     
\end{abstract}

 
\maketitle 

\section{Introduction
\label{sec1} }

A central problem in the field of magneto- or spin--electronics
\cite{Zutic2004} is the spin-injection from a metal into a semiconductor \cite{Olaf}.
In principle it is possible to achieve 100\% spin-polarized injected current
if the magnetic lead is a  half-metallic  material. These compounds are ferromagnets 
where there is a band gap at the Fermi level ($E_F$) for the minority spin
band while the majority spin band is metallic. In such a compound the behavior of 
the interface between the half-metal and the semiconductor is of great importance 
since interface states can kill the half-metallicity.
Although from  point of view of transport a single interface  state 
does not affect the magnetoconductance since the wavefunction is orthogonal to all 
bulk states incident to the interface, its interaction  with other defect  states  
makes the interface states conducting.    

NiMnSb, a member of the Heusler alloys, was the first material to be predicted 
to be a half-metal in 1983 by de Groot and his collaborators \cite{groot}. 
There exist several other ab-initio calculations
on NiMnSb reproducing the results of de Groot \cite{bulkcalc} and Galanakis
\textit{et al.} showed that the gap arises from the hybridization between the $d$ 
orbitals of the Ni and Mn atoms \cite{iosifHalf}. 
Its half-metallicity seems to be well-established experimentally in the case of single crystals
 \cite{Kir-Hans}. Also the so-called full-Heusler alloys like Co$_2$MnGe or Co$_2$CrAl
were predicted to be half-metals \cite{bulkcalc2} and the gap in the case of these
materials arises from states located exclusively at the Co states which are non-bonding
with respect to the other atoms \cite{iosifFull}.

Although films of both half- and full-Heusler alloys attracted a lot of experimental 
attention \cite{Molenkamp,001exper,001exper2}, theoretical calculations 
for the interfaces of these materials with the semiconductors are few. 
All ab-initio results agree that half-metallicity is lost at the interface between
the Heusler alloy and the semiconductor \cite{groot2,Debern,PicozziInter}  but the
interface dependence of the spin-polarization has not been studied in detail.  
Even if half-metallicity is lost it is possible that a high degree of
spin polarization remains at the interface, as it will be shown in this 
contribution, and these structures remain attractive
for realistic applications.

In this communication I study the (001) interfaces of the
half-metallic Co$_2$CrAl Heusler alloy with InP. This 
Heusler alloy has the same experimental lattice constant with the InP within
1\%. 
I take into account all possible interfaces and show that in
all cases a high degree of spin-polarization remains at the interface.
In section \ref{sec2} I discuss the structure of the interface and the 
computational details and in section \ref{sec3} I present and analyze my results.
Finally in section \ref{sec4} I summarize and conclude.

\section{Computational method and structure
\label{sec2} }

In the calculations I used the the full-potential version of the screened
Korringa-Kohn-Rostoker (KKR) Green's function method
\cite{Pap02}
in conjunction with the local spin-density approximation \cite{vosko}
for the exchange-correlation potential \cite{Kohn}.  
The results of Picozzi \textit{et al.} \cite{PicozziInter} and Debernardi \textit{et al.}
\cite{Debern} have shown that atomic positions scarcely change at the interface
and the dominant effect is the expansion or the contraction of the lattice 
along the growth axis to account for the in-plane change of the lattice parameter.
In the case of the interfaces presented here the compounds have similar lattice 
parameters and thus perfect epitaxy can be assumed.
To simulate the interface I used a multilayer consisted of 
15 layers of the half-metal and 9 semiconductor layers.  This
thickness is enough so that the layers in the middle of both the 
half-metallic part and the semiconducting one exhibit
bulk properties. 
I have also converged the  \textbf{k}-space grid, the number of energy 
points and
the tight binding cluster so that the properties of the interfaces do not change
(similar DOS  and spin moments).  
So I  have used a
30$\times$30$\times$4 \textbf{k}-space grid to
perform the integrations in the first Brillouin zone. To evaluate
the charge density one has to integrate the Green's function over an
energy contour in the complex energy plane; for this 42 energy
points were needed. 
A tight-binding cluster of 65 atoms was used in the calculation of
the screened KKR structure constants \cite{zeller97}.  
Finally for the wavefunctions I took angular momentum up to
 $\ell_{max}=3$ into account  and for the
charge density and potential up to $\ell_{max}=6$. 

\begin{figure}
  \begin{center}
\includegraphics[scale=0.6]{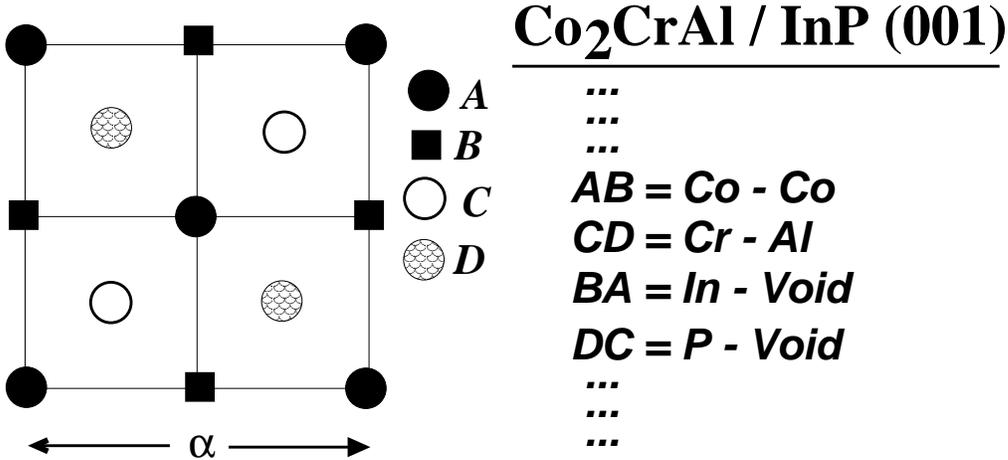}
 \end{center}
\caption{\label{fig1}
Schematic representation of the (001) interface between Co$_2$CrAl and InP.
There are several different combinations at the interface which can be either 
Co/In, Co/P, CrAl/In (shown in the figure) or CrAl/P. 
Note that  there are two inequivalent cobalt atoms at the interface layer or the
subinterface layer. One is sitting at the ``bridge'' site, continuing the zinc-blende
structure of the semiconductor,  and the other at the
``antibridge'' site.}
\end{figure}         

Co$_2$CrAl crystallizes in the  $L2_1$ structure.
The structure of the interface is shown in figure \ref{fig1}.
$L2_1$ structure is similar to the zinc-blende structure and thus perfect epitaxy at
the interface can be considered. There are several combinations at 
the interface, \textit{e.g.} at the Co$_2$CrAl/InP contact the interface can be 
either a Co/In one, Co/P, CrAl/In or CrAl/P. I will keep this definition
through out the paper to denote different interfaces.
Finally I should mention that since my multilayer contains 
15 half-metal and 9 semiconductor layers,
I have two equivalent surfaces at both sides of the half-metallic spacer.

\section{Results and discussion
\label{sec3} }

Interfaces with respect to simple surfaces are more complex systems due
to the hybridization between the orbitals of the atoms of the metallic alloy and the 
semiconductor at the interface. Thus results obtained for the surfaces 
as the ones in reference \cite{iosifSurf} cannot be easily generalized 
for interfaces since for different semiconductors different phenomena can occur.
In Heusler alloys (001) surfaces  the appearance of surface states 
kills the half-metallicity \cite{iosifSurf} but there are cases
like the CrAl-terminated (001)   surface of Co$_2$CrAl where spin-polarization
is as high as 84\%. The case of the multilayers between the 
half-metallic zinc-blende CrAs or CrSe compounds and binary semiconductors is simpler
since for these interfaces the large enhancement of the Cr spin moment kills the 
interface states \cite{iosifZB}.

\subsection{CrAl/In and CrAl/P interfaces 
\label{sec3-1} }

\begin{figure}
  \begin{center}
\includegraphics[scale=0.6]{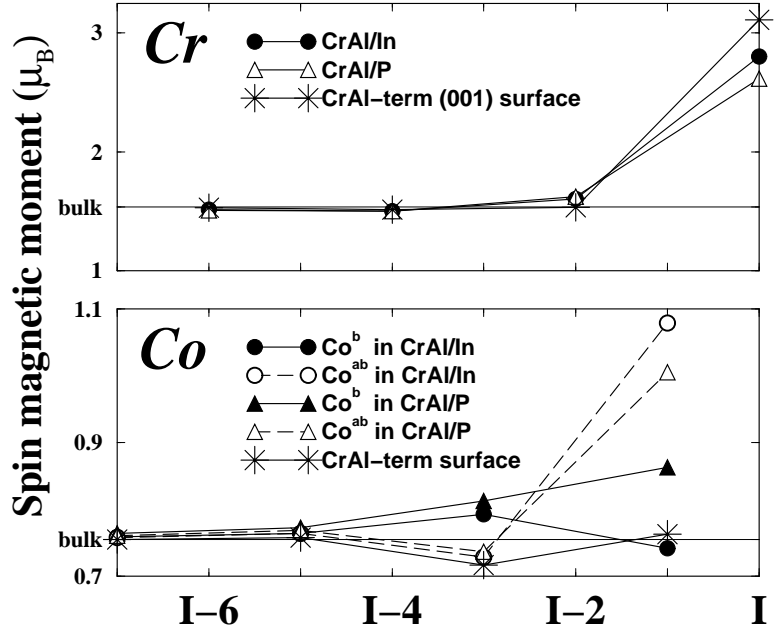}
  \end{center}
\caption{ \label{fig2}
Atom-resolved spin moments in $\mu_B$ for Cr  at the interface (I) and Co at the subinterface (I-1) layer 
and their variation in the film compared 
with the CrAl-terminated  (001) surface. 
Co atoms can sit either at a ``bridge'' site (Co$^\mathrm{b}$) or an ``antibridge'' site (Co$^\mathrm{ab}$).
With the straight horizontal line the bulk values.}
\end{figure}

\begin{figure}
  \begin{center}
\includegraphics[scale=0.6]{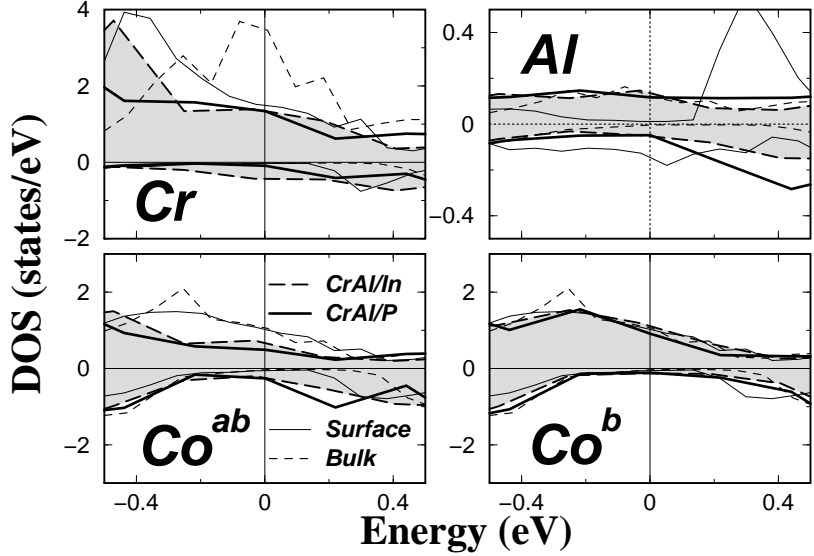}
  \end{center}
\caption{ \label{fig3}
Spin and atom-resolved DOS for the Cr and Al atoms at the interface with In (long dashed line
filled with grey) or P (thick solid line) and the Co atoms at the subinterface layer. 
With solid line the (001) CrAl surface and with the dashed line the bulk results from 
references \cite{iosifSurf} and \cite{iosifFull}, respectively. The zero of the
 energy is chosen to
correspond to the Fermi level. Positive values of the DOS correspond to the majority spin
and negative to the minority.}
\end{figure}

Firstly  I will concentrate my study on the 
case of the CrAl-terminated Co$_2$CrAl(001) film. In a previous article (see reference 
\cite{iosifSurf}) I had  shown that the CrAl (001) terminated  surface was showing
a very high degree of spin-polarization compared to all other surfaces. The mechanism was
quite simple: Cr was loosing 4 out of the 8 first neighboring Co atoms and regained
the charge it was giving away to cobalts in the bulk case. Most of this charge filled
up Cr majority states (in figure \ref{fig3} the majority peak at the Fermi level moves lower in energy) 
and its spin moment was strongly enhanced and due to the stronger 
exchange splitting at the surface the unoccupied Cr states were pushed higher in energy and only the 
surface state due to the Al atoms survived. Actually a similar phenomenon happens at the interface but now 
the increase of the spin moment is smaller since Cr $d$-orbitals hybridise also with 
the In or P $p$-states at the interface. This is clearly seen in figure \ref{fig2} where 
I have gathered the spin moments for the Cr and Co atoms for both interfaces with In and P.
Cr spin moments at the interface are enhanced and reach 2.8$\mu_B$ in the case of the 
interface with In and 2.6$\mu_B$ in the case of the P interface as compared with the 3.1$\mu_B$
of the Cr in the CrAl-terminated surface. The Cr atoms deeper in the half-metallic spacer have 
bulklike spin moments. In the case of the Co atoms the situation is more complicated.
There are two inequivalent Co atoms: the one  at the ``bridge'' site (Co$^\mathrm{b}$) and the one
at the ``antibridge'' site (Co$^\mathrm{ab}$). At the subinterface layer in general 
Co spin moments are strongly enhanced and the moments are larger for the Co atoms at 
the ``antibridge'' sites. If I  add the spin moments of both inequivalent Co's
I  notice that the sum is the same for both the CrAl/In and CrAl/P interfaces and around 2$\mu_B$.
If I  take into account the band structure analysis for the bulk Co$_2$CrAl presented in reference
\cite{iosifFull} that means that both majority $e_u$ states are occupied leading to 
a total Co spin moment of 2 $\mu_B$, while these non-bonding states are unoccupied 
for the minority band. In the case of the I-3 layer the average Co spin moment is equal to
the bulk one and deeper in the film one  finds again the bulk values.

The next question which arises is if this enhancement of the spin moment of Cr is enough to guarantee
a high degree of spin polarization. In figure \ref{fig3} I have plotted  the DOS for the
Cr and Al atoms at the interface and the Co atoms at the subinterface layer for both 
CrAl/In (dashed line filled with grey) and CrAl/P (thick solid line) contacts with respect to 
the surface (solid line) and bulk calculations (dashed line). At the Cr site the spin-polarization is 
almost 100\% for the CrAl/P case and there is a small DOS for the CrAl/In. For the other three atoms
the differences are small between the two different interfaces. Al atom shows a much higher spin polarization
at the Fermi level with respect to the surface results while Co$^\mathrm{ab}$ shows the inverse 
behavior. Notice that the scale along the  DOS axis for the Al atom is different than for the other three.
To make all this more clear, in table \ref{table1} I have gathered the density of states
at the Fermi level for all atoms at the interface for both CrAl/In and CrAl/P interfaces 
together with the results for the CrAl surface. Cobalt has a different behavior depending on
which site it sits at and the ones at the ``bridge'' site behave like in the surface showing  a higher
spin polarization. As already mentioned the Cr spin-polarization is higher for the 
case of the contact with P than with In. In the semiconductor film the only noticeable effect is 
when the In atom is at the interface and it has  a large negative spin-polarization while 
when it sits at the subinterface layer in the case of the CrAl/P contact its net spin-polarization
is almost zero. In total the CrAl/In interface shows a spin polarization of 63\% and the CrAl/P of
65\% as compared to the 84\% of the CrAl surface case. This means that in both interfaces more than 80\%
of the electrons at the Fermi level are of majority spin character and the interface holds a very high degree
of spin-polarization.

\begin{table}
\caption{Number of states at the Fermi level in states/eV units for the atoms 
at the interface for the case of the CrAl/In and CrAl/P interfaces as ratios between 
majority ($\uparrow$) and minority ($\downarrow$) spins together with the
results for the CrAl-terminated (001) surfaces. The last line is the 
spin-polarization $P$ taking into account the interface layers and 
the subinterface ones. \label{table1} }
\begin{indented} 
 \item[]
\begin{tabular}{r|r|r|r} \br
&CrAl/In& CrAl/P & CrAl-surf \\  \mr
Co$^{b}$ ($\uparrow/ \downarrow$)      & 1.18/0.13 & 0.91/0.10 & 1.03/0.06 \\ 
Co$^{ab}$ ($\uparrow/ \downarrow$)     & 0.72/0.12 & 0.49/0.26 & 1.03/0.06 \\ 
Cr ($\uparrow/ \downarrow$)            & 1.39/0.43 & 1.34/0.09 & 1.48/0.03 \\ 
Al ($\uparrow/ \downarrow$)            & 0.15/0.05 & 0.12/0.05 & 0.01/0.15 \\ 
In ($\uparrow/ \downarrow$)            & 0.09/0.25 & 0.07/0.06 & -- \\
Void ($\uparrow/ \downarrow$)          & 0.09/0.08 & 0.02/0.01 & --\\
P  ($\uparrow/ \downarrow$)            & 0.15/0.08 & 0.11/0.08 & --\\
Void ($\uparrow/ \downarrow$)          & 0.02/0.10 & 0.07/0.02 & --\\ \mr
$P$ ($\uparrow - \downarrow \over \uparrow + \downarrow$) & 63\%  & 65\%  & 84\%  \\ \br
\end{tabular}
\end{indented} 
\end{table}

\subsection{Co/In and Co/P interfaces 
\label{sec3-2} }

\begin{figure}
  \begin{center}
\includegraphics[scale=0.6]{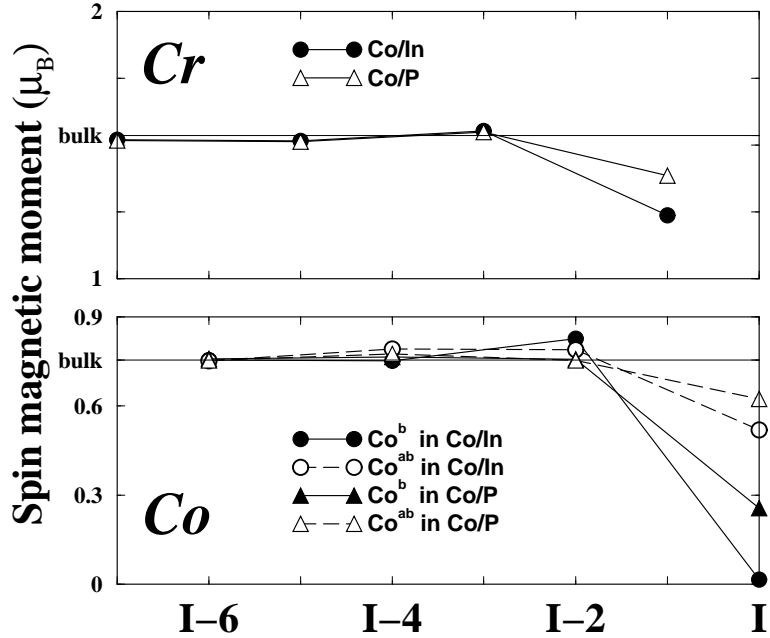}
  \end{center}
\caption{ \label{fig4}
Atom-resolved spin moments in $\mu_B$ for Co  at the interface (I) and Cr at the subinterface (I-1) layer 
and their variation in the spacer. 
Co atoms can sit either at a ``bridge'' site (Co$^\mathrm{b}$) or an ``antibridge'' site (Co$^\mathrm{ab}$).
With the straight horizontal line the bulk values.}
\end{figure}

In the second part of my study I will discuss the case of the interfaces made up by Co
and either an In or a P layer. In figure \ref{fig4} I have gathered the atomic 
spin moments for the Co atoms at the interface and the Cr atoms at the subinterface 
layers and their variation in the film. Spin moments at the interface are strongly 
reduced especially for the Co atoms sitting at the ideal zinc-blende positions,
the so-called ``bridge'' site. The Co$^\mathrm{b}$ spin moment decreases down to $\sim$0.3$\mu_B$ for 
the Co/P interface and the quenching of the Co$^\mathrm{b}$ spin moment is almost 
complete in the case of the Co/In interface. On the other hand the Co$^\mathrm{ab}$ atoms
show a more modest decrease of their spin moment by $\sim 0.15-0.2\mu_B$ with respect
to the bulk value denoted by a straight line in the figure. The Cr atoms at the 
interface layer (I-1) follow through hybridization the behavior of the Co spin moments
and their spin moment is  $\sim 0.15-0.25\mu_B$ smaller than the bulk value.
As soon as one reaches the second layer below the interface, atoms regain a bulklike 
behavior and moments are close to their bulk values.

The behavior of the Co spin moments at the interface has been also observed in the 
case of the Co$_2$MnGe/GaAs contacts studied by Picozzi and collaborators 
\cite{PicozziInter}. For this compound Co in the bulk has a spin moment of $\sim1\mu_B$
but at the Co/Ga or Co/As interfaces the decrease of the Co$^\mathrm{b}$ is as much as
0.8$\mu_B$ while for Co$^\mathrm{ab}$ atoms the reduction of the spin moment is 
only 0.2$\mu_B$. It seems that the reduction of the Co spin moment depends strongly
on the hybridization between the Co $d$-orbitals and the $p$-orbitals of the 
semiconductor. Already for Co at the ``bridge'' site the orbitals hybridise much stronger
than in the case of the Co at the ``antibridge'' site resulting in a larger 
decrease of the spin moment. Also in the system which I study
hybridization is much more important in the case of an In interface layers than of a 
P one leading to the complete quenching of the Co$^\mathrm{b}$ spin moment.
Similar results have been obtained in the case of an Fe film capped by GaAs 
\cite{Scheffler}. In this case an ad layer of Ga or As on top of the Fe film suppresses 
the Fe magnetic moments, the effect being particularly pronounced in As-capped case, 
due to the stronger covalent bonding between the As  and the Fe atoms.

\begin{figure}
  \begin{center}
\includegraphics[scale=0.6]{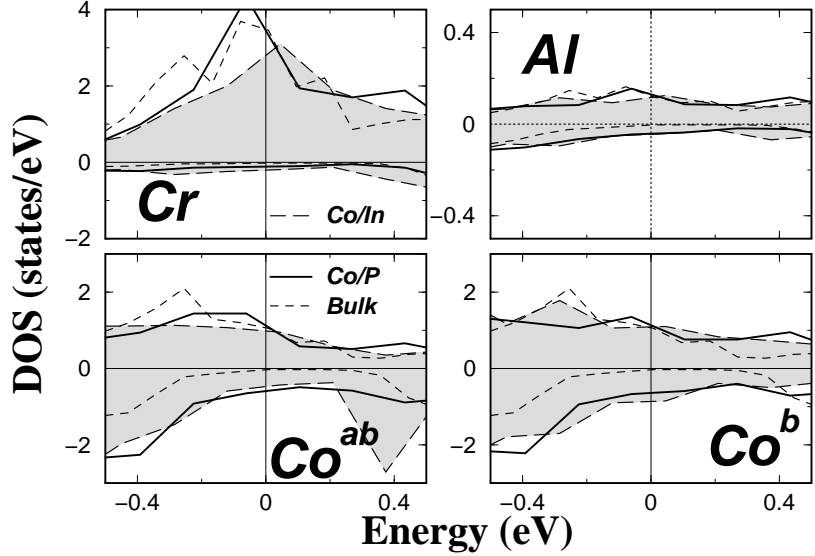}
  \end{center}
\caption{ \label{fig5}
Spin and atom-resolved DOS for the Co atoms  at the interface with In (long dashed line
filled with grey) or P (thick solid line) and the Cr and Al atoms  at the subinterface layer. 
With the dashed line the bulk results from 
reference \cite{iosifFull}.}
\end{figure}

Finally I will discuss the atom-resolved DOS at the interface.
The hybridization of the Co $d$-states with the $p$ states of either In or P  at the 
interface not only reduces the spin moment but also kills half-metallicity at the Co 
sites as can be seen for both Co$^\mathrm{b}$ and Co$^\mathrm{ab}$ in figure \ref{fig5}.
Cr and Al atoms at the subinterface layer have an environment very similar to the bulk
case and although the spin moment of Cr is slightly decreased, as I have already 
mentioned above, its DOS remains similar to the bulk one and it keeps a 
very high majority DOS at the Fermi level similar to the bulk DOS. This effect
largely compensates the loss of half-metallicity of the Co atoms and 
the spin-polarization at the Fermi level, if I take into account the layers close to
the interface (as in table \ref{table1}), is $\sim$56\% for the Co/In interface and
$\sim$74\% for the Co/P interface. Thus 78\% of the electrons at the Fermi level 
for the Co/In interface and 87\% for the Co/P one are of majority character.

In the case of the CrAl interfaces the high spin-polarization was due to the 
large enhancement of the Cr spin moment which weakened the effect of interface states 
although the Cr majority DOS at the Fermi level was considerably smaller than in 
the bulk case; the peak moved lower in energy to accommodate the extra electrons (see figure \ref{fig3}).
In the case of the Co interfaces, although Co themselves present almost a zero net
spin-polarization at the Fermi level, Cr atoms in the subinterface layer keep
the high majority DOS of the bulk  (see figure \ref{fig5}) and the resulting spin-polarization is similar to the 
CrAl interfaces.

\section{Summary and conclusions 
\label{sec4} }

I have studied  the electronic and magnetic properties
of the (001) interfaces between the half-metal Co$_2$CrAl and the binary semiconductor InP 
using a full-potential  ab-initio technique. 
 When the interface 
is made up from a CrAl layer then the Cr spin moment is strongly enhanced at the interface as 
was the case for the CrAl-terminated (001) surfaces. This enhancement limits the effect
of the interface states and in both type of contacts (In or P as interface layer) the interface presents
a very high spin-polarization of $\sim$63-65\%, thus more than 80\%
of the electrons at the Fermi level are of majority spin character. 
On the other hand interfaces made up by Co layers present a 
large decrease of the Co spin moments but, due the bulklike density of
states of the Cr atoms in the subinterface layer, they keep a high degree of 
spin-polarization: 56\% for the Co/In interface and 74\% for the Co/P one .

Interface states are important because their interaction with defects makes them conducting and 
lowers the efficiency of devices based on spin-injection. Thus building up interfaces with the highest 
spin-polarization possible like the ones proposed here is a perquisite but not a guarantee to get highly spin-polarized 
current in spin-injection experiments.


\section*{References} 

\begin{thebibliography}{99}

\bibitem{Zutic2004}
\v{Z}uti\'c I,  Fabian J and  Das Sarma S 2004 \RMP 
 \textbf{76} 323

\bibitem{Olaf}
Wunnicke O,  Mavropoulos Ph,  Zeller R and Dederichs P H 2004 \JPCM
\textbf{16} 4643;
Wunnicke O,  Mavropoulos Ph,  Zeller R, Dederichs P H and  Gr\"undler D
2002 \PR B \textbf{65} 241306; 
Mavropoulos Ph,  Wunnicke O and  Dederichs P H 2002 \PR B \textbf{66}
024416

\bibitem{groot}
de Groot R A, Mueller F M, van Engen P G and Buschow K H J 1983 
\PRL \textbf{50} 2024

\bibitem{bulkcalc}
Kulatov E and Mazin II 1990 \JPCM \textbf{2} 
343; Halilov S V and  Kulatov E T  1991 \JPCM \textbf{3} 
6363;  Wang X,  Antropov V P and 
Harmon B N 1994 \textit{IEEE Trans. Magn.} \textbf{30} 4458;
Youn S J and Min B I 1995 \PR B  \textbf{51} 10436;  Antonov V N,
Oppeneer P M,  Yaresko A N,  Perlov A Ya and Kraft Y 1997 
\PR B \textbf{56} 13012; 
Galanakis I, Ostanin S, Alouani M, Dreyss\'e H and  Wills J M
(2000) \PR B \textbf{61} 4093     

\bibitem{iosifHalf}
Galanakis I, Dederichs P H and Papanikolaou N  2002 \PR B 
 \textbf{66} 134428 

\bibitem{Kir-Hans}
Kirillova M N, Makhnev A A, Shreder E I, Dyakina V P and Gorina N
B 1995 \PSS (b)  \textbf{187} 231;
Hanssen K E H M and Mijnarends P E 1990 \PR B \textbf{34} 5009;
Hanssen K E H M, Mijnarends P E, Rabou L P L M and  Buschow
K H J 1990 \PR B \textbf{42} 1533   

\bibitem{bulkcalc2}
Miura Y,  Nagao K and  Shirai M 2004 \PR B \textbf{69} 144113;
Picozzi S, Continenza A and  Freeman A J 2003 \PR B \textbf{66} 094421; 
Ishida S, Fujii S,  Kashiwagi S
and  Asano S 1995 \JPSJ \textbf{64} 2152


\bibitem{iosifFull}
 Galanakis I,  Dederichs P H and Papanikolaou N 2002 \PR B 
 \textbf{66} 174429;  Galanakis I 2004 \JPCM  \textbf{16} 3089

\bibitem{Molenkamp}
Bach P, Bader A S,  R\"uster C, Gould C,  Becker C R,  Schmidt G,
Molenkamp L W,  Weigand W,  Kumpf C,  Umbach E,  Urban R, 
Woltersdorf G and  Heinrich B 2003 \textit{Appl. Phys. Lett.}  
\textbf{83} 521; Bach P, R\"uster C, Gould C,  Becker C R,  Schmidt G and
Molenkamp L W 2003 \textit{J. Cryst. Growth} \textbf{251} 323

\bibitem{001exper}
van Roy W, Wojcik M, Jedryka E, Nadolski S,  Jalabert D, 
Brijs B, Borghs G and De Boeck J 2003 \textit{Appl. Phys. Lett.} \textbf{83}
4214;
van Roy W, de Boeck J, Brijs B and Borghs G 2000 \textit{Appl.
Phys. Lett.} \textbf{77} 4190;
Schlomka J P, Tolan M and  Press W 2000 \textit{Appl. Phys. Lett.}
\textbf{76} 2005;
Ristoiu D, Nozi\`eres J P, Borca C N, Komesu T, Jeong H -K and
Dowben P A 2000 \textit{Europhys. Lett.} \textbf{49} 624; Ristoiu
D, Nozi\`eres J P, Borca C N, Borca B and Dowben P A 2000
\textit{Appl. Phys. Lett.}  \textbf{76} 2349;
Giapintzakis J,  Grigorescu C, Klini A,  Manousaki A,
Zorba V, Androulakis J,  Viskadourakis Z and Fotakis C
2002 \textit{Appl. Phys. Lett.} \textbf{80} 2716


\bibitem{001exper2}
Yang F Y, Shang C H, Chien C L, Ambrose T,  Krebs J J, 
Prinz G A, Nikitenko V I, Gornakov V S,  Shapiro A J and 
Shull R D 2002 \PR B \textbf{65} 174410; 
Ambrose T,  Krebs J J and Prinz G A 2000 \textit{Appl. Phys. Lett.}
\textbf{76} 3280;
Raphael M P, Ravel B, Willard M A,  Cheng S F, Das B N,  Stroud R M, Bussmann K M,
Claassen J H and  Harris V G 2001 \textit{Appl. Phys. Lett.} \textbf{70} 4396;
Chen Y J, Basiaga D, O'Brien J R and Heiman D 2004
\textit{Appl. Phys. Lett.} \textbf{84} 4301;
Elmers H J, Fecher G H,  Valdaitsev D, Nepijko S A,  Gloskovskii A,  Jakob G,
Schonhense G,  Wurmehl S,  Block T,  Felser C,  Hsu P C,
Tsai W L and Cramm S 2003 \PR B \textbf{67} 104412
 
\bibitem{groot2}
Wijs G A and  de Groot R A 2001 \PR B  \textbf{64} R020402 

\bibitem{Debern}
 Debernardi A, Peressi M and Baldereshi A 2003 
\textit{Mat. Sci. Eng.} C \textbf{23} 743 

\bibitem{PicozziInter}
Picozzi S, Continenza A and Freeman A J 2003 \textit{J. Phys. Chem. Solids} 
\textbf{64} 1697; ibid 2003 \JAP \textbf{94} 4723

\bibitem{zeller95}
Zeller R, Dederichs P H, \'Ujfalussy B, Szunyogh L and Weinberger
P 1995 \PR B  \textbf{52} 8807   

\bibitem{Pap02}
Papanikolaou N, Zeller R and Dederichs P H 2002  \JPCM \textbf{14} 
2799 

\bibitem{vosko}
Vosko S H, Wilk L and Nusair N 1980 \CJP \textbf{58} 1200 

\bibitem{Kohn}
Hohenberg P and Kohn W 1964 \PR  \textbf{136} B864;
Kohn W and Sham L J 1965 \PR \textbf{140} A1133

\bibitem{zeller97}
Zeller R 1997 \PR B \textbf{55} 9400 

\bibitem{iosifZB}
Mavropoulos Ph,  Galanakis I and Dederichs P H 2004 \JPCM
\textbf{16} 4261 

\bibitem{iosifSurf}
Galanakis I 2002 \JPCM \textbf{14}   6329

\bibitem{Scheffler}
Erwin S C,  Lee S H and Scheffler M (2002) \PR B \textbf{65}  205422

\end{thebibliography}
\end{document}